\documentclass[10pt,twocolumn, conference]{IEEEtran}

\usepackage{amsmath}
\usepackage{amssymb}
\usepackage[dvips]{graphicx}
\usepackage{epsfig}

\newcommand{\figsize}{0.31}

\begin{document}
\title{Pilot-Symbol-Assisted Communications with Noncausal and Causal Wiener Filters}
\author{\authorblockN{Sami Akin \hspace{1.0cm}
Mustafa Cenk Gursoy}
\authorblockA{Department of Electrical Engineering\\
University of Nebcanonicalnew1raska-Lincoln\\ Lincoln, NE 68588\\ Email:
sakin1@bigred.unl.edu, gursoy@engr.unl.edu}}
\date{}

\maketitle

\begin{abstract}
In this paper, pilot-assisted transmission over time-selective flat
fading channels is studied. It is assumed that noncausal and causal
Wiener filters are employed at the receiver to perform channel
estimation with the aid of training symbols sent periodically by the
transmitter. For both filters, the variances of estimate errors are
obtained from the Doppler power spectrum of the channel.
Subsequently, achievable rate expressions are provided. The training
period, and data and training power allocations are jointly
optimized by maximizing the achievable rate expressions. Numerical
results are obtained by modeling the fading as a Gauss-Markov
process. The achievable rates of causal and noncausal filtering
approaches are compared. For the particular ranges of parameters
considered in the paper, the performance loss incurred by using a
causal filter as opposed to a noncausal filter is shown to be small.
The impact of aliasing that occurs in the undersampled version of
the channel Doppler spectrum due to fast fading is analyzed.
Finally, energy-per-bit requirements are investigated in the
presence of noncausal and causal Wiener filters.

\end{abstract}
\vspace{-0.25cm}
\section{Introduction}
In wireless communications, channel conditions vary over time due to
mobility and changing environment. If the channel conditions are not
known a priori, practical wireless systems generally employ training
sequences to perform channel estimation, receiver adaptation and
optimal decoding \cite{Tong}. Cavers in \cite{caver1} and
\cite{caver2} conducted one of the early studies in this area and
provided an analytical approach to the design of pilot-assisted
transmissions. Recently, there has been much interest in the
optimization of training parameters using an information-theoretic
approach. Hassibi and Hochwald \cite{Hassibi} considered the
multiple-antenna Rayleigh block fading channel and optimized the
power and duration of training signals by maximizing a capacity
lower bound. Adirredy \emph{et al.} \cite{adireddy} investigated the
optimal placement of pilot symbols and showed that the periodical
placement maximizes the data rates. In general, the amount,
placement, and fraction of pilot symbols in the data stream have
considerable impact on the achievable data rates.

Considering adaptive coding of data symbols without feedback to the
transmitter, Abou-Faycal \emph{et al.} \cite{faycal2} studied the
data rates achieved with pilot-symbol-assisted modulation (PSAM)
over Gauss-Markov channels. The authors in \cite{faycal3} also
studied the PSAM over Gauss-Markov channels and analyzed the power
allocation of data symbols when the pilot symbol has fixed power.
They showed that the power has a decreasing character with respect
to the distance to the pilot symbol. In similar settings,
\cite{Gursoy2} analyzed the training power when the data power is
distributed uniformly. More recently, we in \cite{Gursoy3} jointly
optimized the pilot symbol period and power allocation among pilot
and data symbols by maximizing the achievable rates in Gauss-Markov
fading channels. Ohno and Giannakis \cite{Giannakis2} considered
general slowly-varying fading processes. Employing a noncausal
Wiener filter for channel estimation at the receiver, they obtained
a capacity lower bound and optimized the spacing of training symbols
and training power. Baltersee \emph{et al.} in \cite{Baltersee1} and
\cite{Baltersee2} have also considered using a noncausal Wiener
filter to obtain a channel estimate, and they optimized the training
parameters by maximizing achievable rates in single and multiple
antenna channels.

In this paper, we study training-based transmission and reception
schemes over a-priori unknown, time-selective Rayleigh fading
channels. Since causal operation is crucial in real-time,
delay-constrained applications, we consider the use of causal, as
well as noncausal, Wiener filters for channel estimation. We
optimize the training parameters by maximizing a capacity lower
bound. Although the treatment is general initially, we concentrate
on the Gauss-Markov channel model for numerical analysis. As another
contribution, we analyze fast fading channels and the impact upon
the performance of aliasing due to under-sampling of the channel.
\vspace{-0.15cm}
\section{Channel Model}
The time-selective Rayleigh channel is modeled as
\vspace{-0.0cm}
\begin{equation}
y_{k}=h_{k}x_{k}+n_{k}  \quad k=1,2,3, \ldots
\end{equation}
where $y_{k}$ is the complex channel output, $x_{k}$ is the complex
channel input, $\{n_k\}$ is assumed to be a sequence of independent
and identically distributed (i.i.d.) zero-mean Gaussian random
variables with variance $\sigma_{n}^{2}$, and $\{h_k\}$ is the
sequence of fading coefficients. $\{h_k\}$ is assumed to be a
zero-mean stationary Gaussian random process with power spectral
density $S_{h}(e^{jw})$. It is further assumed that $x_{k}$ is
independent of $h_{k}$ and $n_{k}$. While both the transmitter and
the receiver know the channel statistics, neither has prior
knowledge of instantaneous realizations of the fading coefficients.
Note that the discrete-time model is obtained by sampling the
received signal every $T_{s}$ seconds.
\vspace{-0.15cm}
\section{Pilot Symbol-Assisted Transmission and Reception}
We consider pilot-assisted transmission where periodically inserted
pilot symbols, known by both the sender and the receiver, are used
to estimate the fading coefficients of the channel using a Wiener
filter. We assume the simple scenario where a single pilot symbol is
transmitted every $M$ symbols while $M-1$ data symbols are
transmitted in between the pilot symbols. We consider the following
average power constraint
\begin{equation}\label{averagepower}
\frac{1}{M}\sum_{k=lM}^{(l+1)M-1}E\left[|x_{k}|^2\right]\leq P\quad
l = 0,1,2,\ldots
\end{equation} on the input.
Therefore, the total average power allocated to the pilot and data
transmission over a duration of $M$ symbols is limited by $MP$.
Communication takes place in two phases. In the training phase, the
transmitter sends pilot symbols and the receiver estimates the
channel coefficients. In this phase, the channel output is given by
\begin{equation}
y_{lM}=h_{lM}\sqrt{P_{t}}+n_{lM}
\end{equation}
where $P_{t}$ is the power allocated to the pilot symbol. In the
data transmission phase, data symbols are transmitted. In this
phase, the input-output relationship can be written as
\begin{equation}\label{input-output}
y_{k}=\widehat{h}_{k}x_{k}+\widetilde{h}_{k}x_{k}+n_{k} \quad
lM<k\leq (l+1)M-1
\end{equation}
where $\widehat{h}_{k}$ and $\widetilde{h}_{k}$ are the estimated
channel coefficient and the error in the estimate at  sample time
$k$, respectively. Note that $\widehat{h}_{k}$ and
$\widetilde{h}_{k}$ for $lM<k\leq (l+1)M-1$ are uncorrelated
zero-mean circularly symmetric complex Gaussian random variables
with variances $\sigma_{\widehat{h}_{k}}^{2}$ and
$\sigma_{\widetilde{h}_{k}}^{2}$, respectively.

\section{Achievable Rates}
For the estimation of the fading coefficients, we assume that a
Wiener filter, which is the optimum linear estimator in the
mean-square sense, is employed at the receiver. Note that since
pilot symbols are sent with a period of $M$, the channel is sampled
every $MT_{s}$ seconds. Therefore we have to consider the
under-sampled version of the channel's Doppler spectrum which is
given by
\begin{equation}\label{undersampled}
S_{h,m}(e^{jw})=\frac{1}{M}\sum_{k=0}^{M-1}e^{jm(w-2\pi
k)/M}S_{h}\left(e^{j(w-2\pi k)/M}\right).
\end{equation}
Also shown in \cite{Giannakis2}, it can easily be seen from
\cite{linear} that the channel MMSE for the noncausal Wiener filter
at time $Ml+m$ is given by
\begin{equation}\label{estimation error}
\sigma_{\widetilde{h}_{Ml+m}}^{2}=\sigma_{h}^{2}-\frac{1}{2\pi}\int_{-\pi}^{\pi}\frac{P_{t}|S_{h,m}(e^{jw})|^{2}}{P_{t}S_{h,0}(e^{jw})+\sigma_{n}^{2}}dw
\end{equation}
where $P_{t}$ again denotes the power allocated to one pilot symbol.
On the other hand, from \cite{linear}, we can also easily find that
the channel MMSE at time $Ml+m$ for the causal Wiener filter is
given by
\begin{align}\label{estimation error causal}
\sigma_{\widetilde{h}_{Ml+m}}^{2} &= \sigma_{h}^{2}
-\frac{1}{2\pi}\int_{-\pi}^{\pi}\frac{P_{t}|S_{h,m}(e^{jw})|^{2}}{P_{t}S_{h,0}(e^{jw})+\sigma_{n}^{2}}dw\nonumber
\\&+\frac{1}{2\pi}\int_{-\pi}^{\pi}\frac{P_{t}}{r_{e}}\left|\left\{\frac{S_{h,m}(e^{jw})}{L^{*}(e^{jw})}\right\}_{-}\right|^{2}dw
\end{align}
where $L^{*}(e^{jw})$ is obtained from the canonical factorization
of the channel output's sampled power spectral density at $m=0$,
which is given by
\begin{equation}\label{canonical}
P_{t}S_{h,0}(e^{jw})+\sigma_{n}^{2}=r_{e}L(e^{jw})L^{*}(e^{jw}).
\end{equation}
The operators $\{\}_{+}$ and $\{\}_{-}$ yield the causal and the
anti-causal part of the function to which they are applied,
respectively. Note that, using the orthogonality principle, we have
\begin{equation}\label{orthogonality}
\sigma_{\widehat{h}_{Ml+m}}^{2}=\sigma_{h}^{2}-\sigma_{\widetilde{h}_{Ml+m}}^{2}
\end{equation}
where $\sigma_{\widehat{h}_{Ml+m}}^{2}$ is the variance of the
channel estimate at time $Ml+m$. Similarly as in \cite{Gursoy3},
treating the error in (\ref{input-output}) as another source of
additive noise and assuming that
\begin{equation}\label{new-error}
w_{k}=\widetilde{h}_{k}x_{k}+n_{k}
\end{equation}
is zero-mean Gaussian noise with variance
\begin{equation}\label{new-variance}
\sigma_{w_{k}}^{2}=\sigma_{\widetilde{h}_{k}}^{2}P_{m}+\sigma_{n}^{2}
\end{equation}
we obtain the following lower bound on the channel capacity:
\begin{equation}\label{lowerboundsimplified}
C\geq\frac{1}{M}\sum_{m=1}^{M-1}E\left\{\log\left(1+\frac{P_{m}\sigma_{\widehat{h}_{m}}^{2}}{P_{m}\sigma_{\widetilde{h}_{m}}^{2}+\sigma_{n}^{2}}|\xi|^{2}\right)\right\}
\end{equation}
where $\xi$ is a zero-mean, unit-variance, circularly symmetric
complex Gaussian random variable and
$P_{m}=E\left[|x_{Ml+m}|^{2}\right]$ denotes the power of the
$m^{th}$ data symbol after the pilot symbol. Note that the error
variance $\sigma_{\widetilde{h}_{Ml+m}}^{2}$ depends in general  on
$m$ and hence the location of the data symbol with respect to the
pilot symbol. However, if the fading slowly varies and the channel
is sampled sufficiently fast, we can satisfy $2 \pi f_D \le \pi/M$
where $f_D$ is the maximum Doppler frequency of the channel. In this
case, $M \le \frac{1}{2f_D}$. We can see from the Nyquist's Theorem
that there is no aliasing in the under-sampled version of the
channel's Doppler spectrum, and hence
$|S_{h,m}(e^{jw})|=|S_{h,0}(e^{jw})|=|S_{h}(e^{jw/M})|/M$, for
$m\in[1,M-1]$ and $-\pi\leq w\leq\pi$. Therefore, (\ref{estimation
error}) reduces to
\begin{align}\label{estimation error simple}
\!\!\!\!\!\!\sigma_{\widetilde{h}_{Ml+m}}^{2}&=\sigma_{h}^{2}-\frac{1}{2\pi}\int_{-\pi}^{\pi}\frac{P_{t}|S_{h,0}(e^{jw})|^{2}}{P_{t}S_{h,0}(e^{jw})+\sigma_{n}^{2}}dw\nonumber
\\&=\sigma_{h}^{2}-\frac{1}{2\pi}\int_{-\pi/M}^{\pi/M}\frac{P_{t}|S_{h}(e^{jw})|^{2}}{P_{t}S_{h}(e^{jw})+M\sigma_{n}^{2}}dw=\sigma_{\widetilde{h}}^{2},
\end{align}
and also (\ref{estimation error causal}) can be expressed as
\begin{align}\label{estimation error causal simple}
\!\!\!\!\!\!\sigma_{\widetilde{h}_{Ml+m}}^{2} &= \sigma_{h}^{2}
-\frac{1}{2\pi}\int_{-\pi}^{\pi}\frac{P_{t}|S_{h,0}(e^{jw})|^{2}}{P_{t}S_{h,0}(e^{jw})+\sigma_{n}^{2}}dw\nonumber
\\&+\frac{1}{2\pi}\int_{-\pi}^{\pi}\frac{P_{t}}{r_{e}}\left|\left\{\frac{S_{h,0}(e^{jw})}{L^{*}(e^{jw})}\right\}_{-}\right|^{2}dw\nonumber
\\
&=\sigma_{h}^{2}
-\frac{1}{2\pi}\int_{-\pi/M}^{\pi/M}\frac{P_{t}|S_{h}(e^{jw})|^{2}}{P_{t}S_{h}(e^{jw})+M\sigma_{n}^{2}}dw\nonumber
\\&+\frac{1}{2\pi}\int_{-\pi/M}^{\pi/M}\frac{P_{t}}{Mr_{f}}\left|\left\{\frac{S_{h}(e^{jw})}{F^{*}(e^{jw})}\right\}_{-}\right|^{2}dw
=\sigma_{\widetilde{h}}^{2},
\end{align}
where
\begin{equation}\label{canonicalnew}
\frac{P_{t}S_{h}(e^{jw})}{M}+\sigma_{n}^{2}=r_{f}F(e^{jw})F^{*}(e^{jw}).
\end{equation}
Therefore, under this assumption, the error variances become
independent of $m$. Since the estimate quality is the same for each
data symbol regardless of its position with respect to the pilot
symbol, uniform power allocation among the data symbols is optimal
and we have
\begin{equation}\label{average power}
P_{m}=\frac{MP-P_{t}}{M-1} = P_0.
\end{equation}

Then, we can rewrite (\ref{lowerboundsimplified}) as
\begin{equation}\label{lowerboundson}
C\geq\frac{M-1}{M}E\left\{\log\left(1+\frac{P_{0}\sigma_{\widehat{h}}^{2}}{P_{0}\sigma_{\widetilde{h}}^{2}+\sigma_{n}^{2}}|\xi|^{2}\right)\right\}.
\end{equation}

\section{Optimizing Training Parameters in Gauss-Markov Channels}
In this section, we assume that the fading process is modeled as a
first-order Gauss-Markov process, whose dynamics is described by
\begin{equation}
h_{k}=\alpha h_{k-1}+z_{k} \quad 0\leq\alpha\leq1\quad
k=1,2,3,\ldots
\end{equation}
where \{$z_{k}$\} are i.i.d. circular complex Gaussian variables
with zero mean and variance equal to (1-$\alpha^{2}) \sigma_{h}^2$.
The power spectral density of the Gauss-Markov process with variance
$\sigma_{h}^{2}$ is given by
\begin{equation}\label{psd}
S_h(e^{jw})=\frac{(1-\alpha^{2})\sigma_{h}^{2}}{1+\alpha^{2}-2\alpha\cos(w)}.
\end{equation}
Note that $S_h(e^{jw})$ in (\ref{psd}) is not bandlimited and hence
the condition $2 \pi f_D \le \pi/M$ can only be satisfied when $M=1$
which is not a viable strategy. However if the fading is
slowly-varying and hence the value of $\alpha$ is close to 1, the
Doppler spectrum $S_h(e^{jw})$ decreases sharply for large
frequencies and most of the energy is accumulated at low Doppler
frequencies. 
We can easily find that the frequency ranges $[-\pi/49,\pi/49]$,
$[-\pi/9,\pi/9]$ and $[-\pi/4,\pi/4]$ contain more than 90 $\%$ of
the power when $\alpha=0.99, 0.95$, and $0.90$, respectively. Hence,
if $M \le 49, 9$, and 4, respectively, in these cases, the impact of
aliasing will be negligible. Otherwise, ignoring the effect of
aliasing will decrease the error variance and hence the achievable
rates under this assumption will be higher than those obtained when
aliasing is considered.

In the Gauss-Markov model, the error variance for the noncausal
Wiener filter can easily be obtained from (\ref{estimation error}).
In order to obtain the error variance for the causal filter in the
absence of aliasing, we have to perform the canonical factorization.
We begin with rewriting (\ref{canonical}) as
\begin{equation}\label{canonicalnew}
\frac{P_{t}S_{h}(e^{jw/M})}{M}+\sigma_{n}^{2}=r_{f}F(e^{jw/M})F^{*}(e^{jw/M})
\end{equation}
where
\begin{equation*}\label{uppercan}
F(e^{jw})=\frac{1-ue^{-jw}}{1-\alpha e^{-jw}}.
\end{equation*}
From (\ref{canonicalnew}), we can deduce that
\begin{equation}\label{canonicalnew1}
c+\sigma_{n}^{2}\alpha(e^{jw/M}+e^{-jw/M})=r_{f}(1+u)+r_{f}u(e^{jw/M}+e^{-jw/M})
\end{equation}
where
\begin{equation*}
c=\frac{P_{t}}{M}(1-\alpha^{2})\sigma_{h}^{2}+(1+\alpha^{2})\sigma_{n}^{2}.
\end{equation*}
From (\ref{canonicalnew1}), we can write
\begin{equation}\label{rf}
r_{f}=\frac{c+\sqrt{c^{2}-4\alpha^{2}\sigma_{n}^{4}}}{2} \quad
\text{and} \quad u=\frac{\alpha\sigma_{n}^{2}}{r_{f}}
\end{equation}
where $0<u<1$ and $r_{f}>0$. After the canonical factorization, we
can write
\begin{align}\label{causalpart}
\frac{S_{h}(e^{jw/M})}{F^{*}(e^{jw/M})}&=\frac{(1-\alpha^{2})\sigma_{h}^{2}}{(1-\alpha
e^{-jw/M})(1-\alpha e^{jw/M})}\frac{1-\alpha\ e^{jw/M}}{1-ue^{jw/M}}
\nonumber \\&=\frac{(1-\alpha^{2})\sigma_{h}^{2}}{(1-\alpha
e^{-jw/M})(1-ue^{jw/M})}\\&
=B\left[\frac{u\alpha}{e^{jw/M}-\alpha}-\frac{1}{e^{jw/M}-1/u}\right]
\end{align}
where
\begin{equation*}
B=-\frac{(1-\alpha^{2})\sigma_{h}^{2}}{u(1-u\alpha)}.
\end{equation*}
The anti-causal part can be written as
\begin{equation}\label{anticausal}
\left\{\frac{S_{h}(e^{jw/M})}{F^{*}(e^{jw/M})}\right\}_{-}=\frac{(1-\alpha^{2})\sigma_{h}^{2}u}{(1-u\alpha)}\frac{e^{jw/M}}{(1-ue^{jw/M})}.
\end{equation}
After making a change of variables, we have
\begin{equation}\label{anticausal1}
\left\{\frac{S_{h}(e^{jw})}{F^{*}(e^{jw})}\right\}_{-}=\frac{(1-\alpha^{2})\sigma_{h}^{2}u}{(1-u\alpha)}\frac{e^{jw}}{(1-ue^{jw})}.
\end{equation}

\section{Numerical Results}
\subsection{Optimal Parameters and Effects of Aliasing}
\begin{figure}
\begin{center}
\includegraphics[width = \figsize\textwidth]{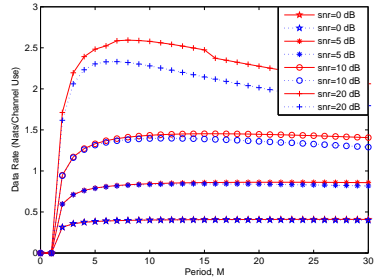}
\caption{Achievable rates when $\alpha=0.99$ for SNR=0, 5,10, and 20
dB. The dotted lines provide rates when aliasing is taken into
account, and the solid lines give the rates when aliasing is
ignored.} \label{fig:fig2}
\end{center}
\end{figure}

In this section, we present our numerical results. Initially, we
consider noncausal Wiener filtering and jointly optimize the
training period, and data and pilot symbol power allocation.
Moreover, we study the effects of aliasing in the under-sampled
channel Doppler spectrum. In Figure \ref{fig:fig2}, we plot the
achievable rates as a function of the training period when
$\alpha=0.99$, i.e., when the channel is changing very slowly, for
SNR values of 0, 5, 10 and 20 dB. In this figure, plotted curves are
obtained with optimal pilot and data power allocation. The dotted
lines give the data rates obtained when aliasing is taken into
account. Solid lines show the rates when aliasing is ignored. As
seen in Fig. \ref{fig:fig2}, when SNR is small, the difference
between the dotted and solid lines is negligible. As SNR increases,
the difference between the lines is also increasing. From this, we
can conceive that the effect of aliasing is also increasing with
increasing power. When $\alpha=0.99$ and aliasing is taken into
account, the optimal training periods are  16, 15, 12 and 7 for SNR
values of 0, 5, 10 and 20 dB, respectively. On the other hand, when
aliasing is ignored, we have optimal values as 25, 21, 16 and 8.
Hence, the optimal training period decreases as SNR increases and
aliasing is considered.

\begin{figure}
\begin{center}
\includegraphics[width = \figsize\textwidth]{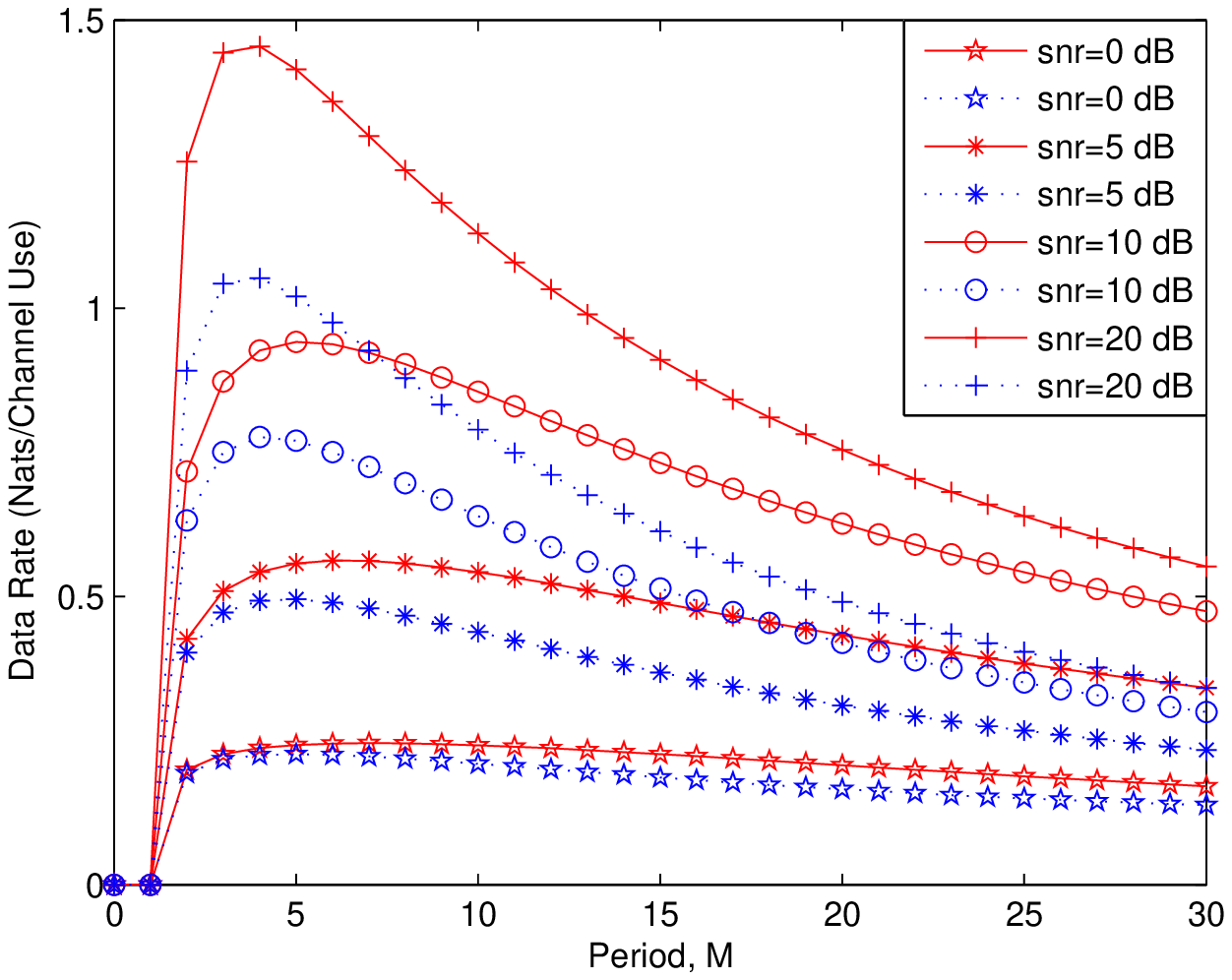}
\caption{Achievable rates when $\alpha=0.90$ for SNR=0, 5,10, and 20
dB. The dotted lines provide rates when aliasing is taken into
account, and the solid lines give the rates when aliasing is
ignored.} \label{fig:fig3}
\end{center}
\end{figure}

In Figure \ref{fig:fig3}, we plot the achievable rates when
$\alpha=0.90$. Comparing Figs. \ref{fig:fig2} and \ref{fig:fig3}, we
observe that aliasing has a more significant impact as $\alpha$
decreases. This is expected since aliasing increases in a faster
changing channel and hence ignoring aliasing provides a looser upper
bound. When $\alpha=0.90$ and aliasing is taken into account, the
optimal training periods are 7, 6, 5 and 4 for SNR values of 0, 5,
10 and 20 dB, respectively. When aliasing is ignored, the optimal
values are 5, 5, 4 and 4, respectively. As before, the optimal
period is decreasing and the effect of aliasing is increasing with
the increasing SNR.

\begin{figure}
\begin{center}
\includegraphics[width = \figsize\textwidth]{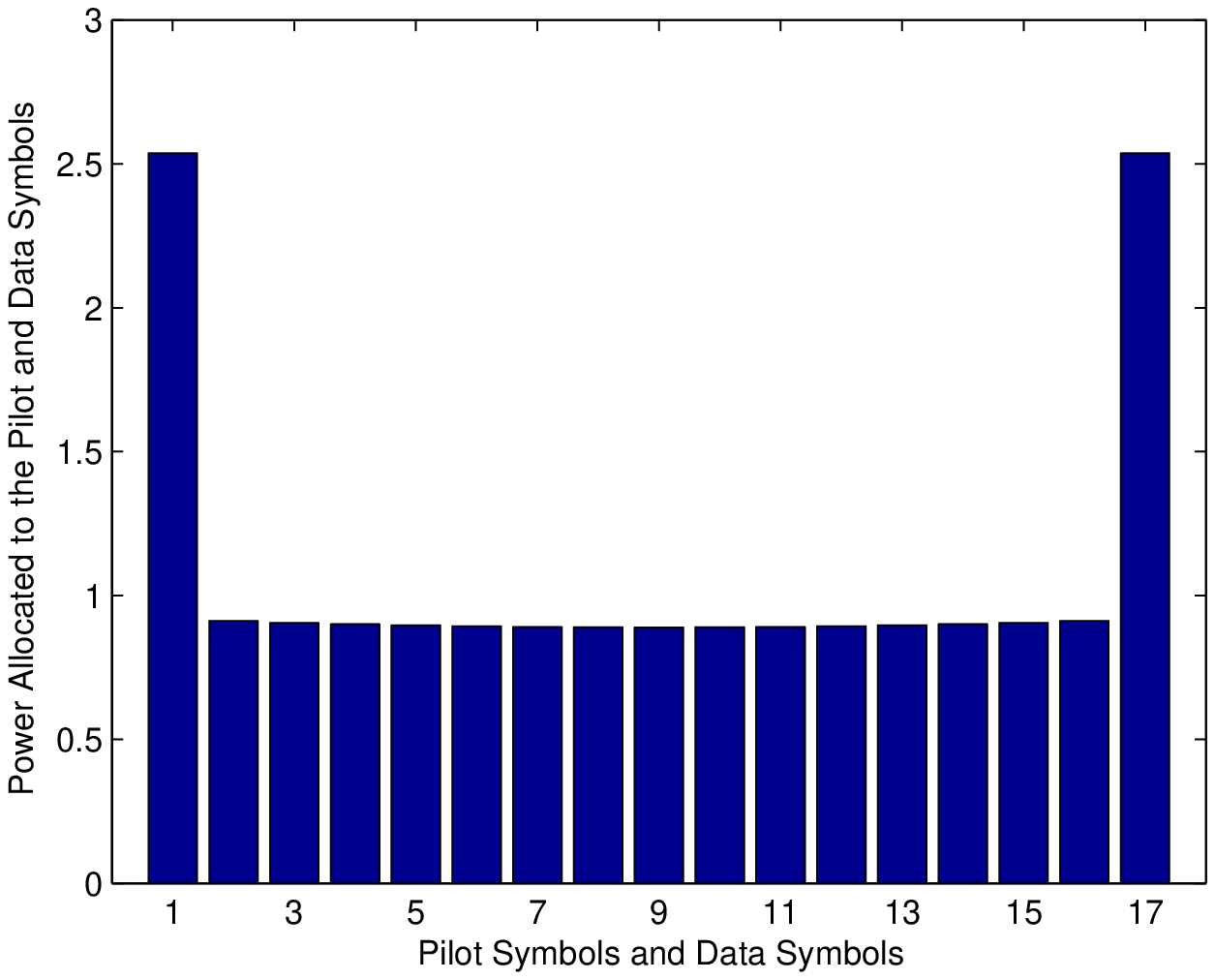}
\caption{The optimal power distribution among the pilot and data
symbols when $\alpha=0.99$ and SNR=0dB. The optimal period is 16.}
\label{fig:fig5}
\end{center}
\end{figure}

\begin{figure}
\begin{center}
\includegraphics[width = \figsize\textwidth]{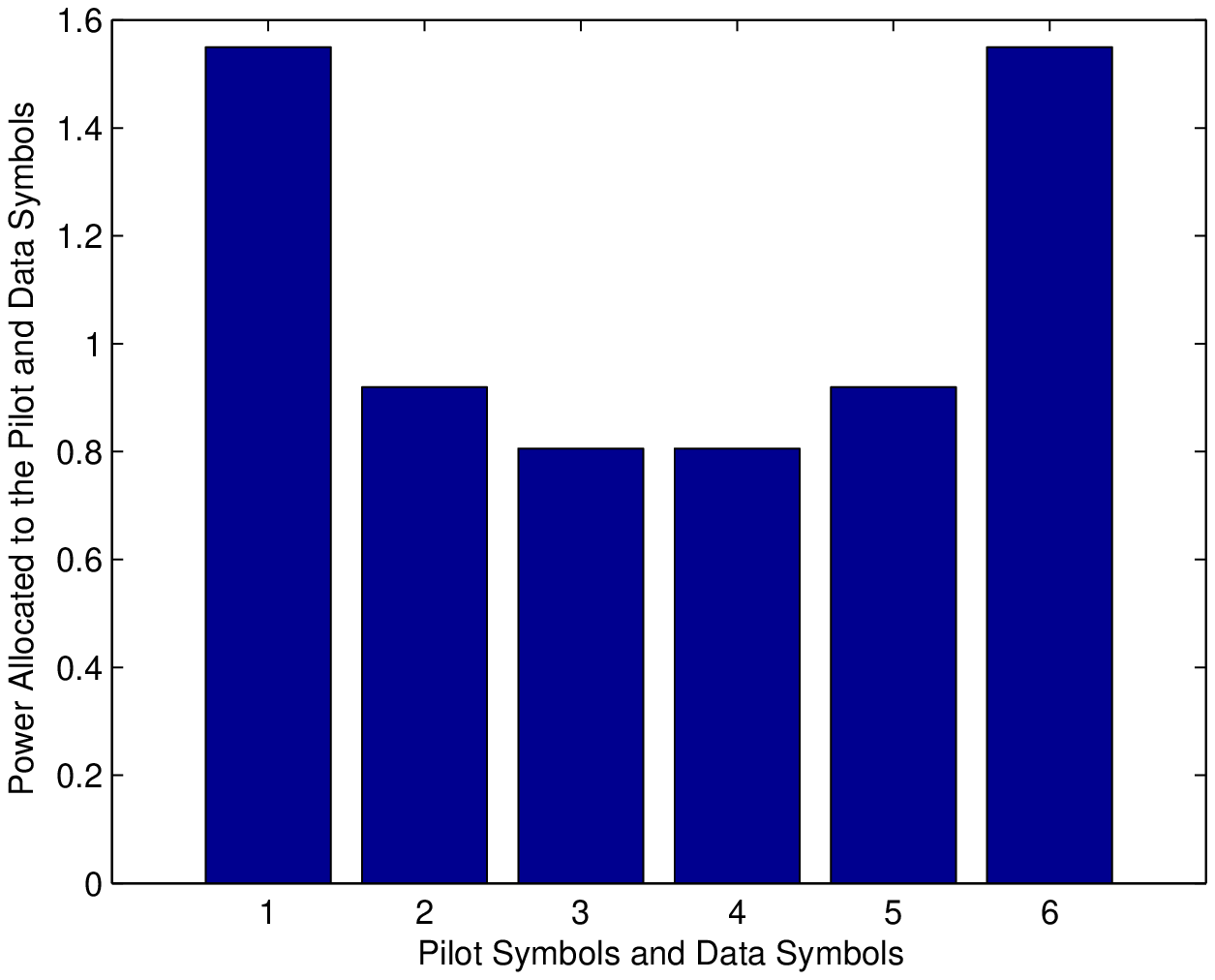}
\caption{The optimal power distribution among the pilot and data
symbols when $\alpha=0.90$ and SNR=0dB. The optimal period is 5.}
\label{fig:fig6}
\end{center}
\end{figure}

Figure \ref{fig:fig5} and Figure \ref{fig:fig6} are the bar graphs
providing the optimal training and data power allocation for
$\alpha=0.99$ and $0.90$, respectively, when the training period is
at its optimal value. In the graphs, the first and the last bars
give the power of the pilot symbols and the ones in between
represent the data symbol power levels. These bar graphs are
obtained when the effect of aliasing on the channel estimation is
taken into account. We can immediately observe from both graphs that
the data symbols farther away from the pilot symbols are allocated
less power because the error in the estimation increases with the
distance to the pilot symbols. In Fig. \ref{fig:fig5}, the decrease
in the allocated power is small since the channel is very slowly
varying and estimate error is almost independent of $m$. On the
other hand, the decrease is more obvious when the channel changes
faster as evidenced in Fig. \ref{fig:fig6}. Furthermore, comparing
Figs. \ref{fig:fig5} and \ref{fig:fig6}, we see that when the
training period value is high, more power is allocated to the pilot
symbol, enabling the system to track the channel more accurately.

\subsection{Causal Filter Performance in the Absence of Aliasing}
\begin{figure}
\begin{center}
\includegraphics[width = \figsize\textwidth]{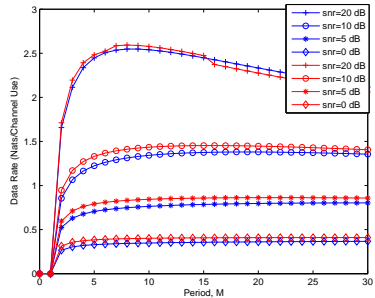}
\caption{Achievable rates vs. training period when noncausal and
causal filters are employed at the receiver. $\alpha=0.99$ and
SNR=0, 5,10, and 20 dB. The solid lines give the rates when a
noncausal filter is used and the dotted lines show the rates when a
causal filter is used.} \label{fig:fig8}
\end{center}
\end{figure}

\begin{figure}
\begin{center}
\includegraphics[width = \figsize\textwidth]{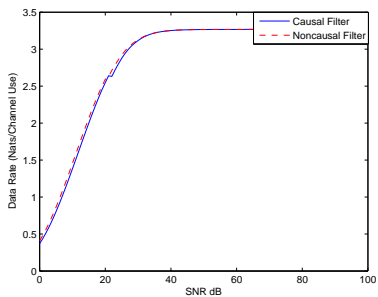}
\caption{Achievable rates vs. SNR when noncausal and causal filters
are employed at the receiver. $\alpha=0.99$. The dashed line gives
the rate when a noncausal filter is used and the solid line shows
the rate when a causal filter is used.} \label{fig:fig9}
\end{center}
\end{figure}

In this section, we study the  performance when a causal Wiener
filter is employed at the receiver. Since it is rather difficult to
obtain the canonical factorization of arbitrary spectrums, we only
consider cases in which the channel is slowly varying and the
aliasing effect can be ignored. In Figure \ref{fig:fig8}, we plot
the achievable rates as a function of the training period for
$\alpha=0.99$ when noncausal and causal Wiener filters are used. We
compare the results when SNR$=0, 5, 10$ and $20$ dB. The dotted
lines provide the rates for the case of the causal filter and the
solid lines show the results for the case of the noncausal filter.
We observe that the optimal training periods are 44, 29, 19 and 9
for the causal filter when SNR$=0, 5, 10$ and $20$, respectively.
For the noncausal filter, the optimal periods are 25, 21, 16 and 8
for the same SNR values. We observe from the plots that the
performance of causal and noncausal filters are very close. In
Figure \ref{fig:fig9}, we plot the achievable rates as a function of
SNR at optimal periods obtained by using causal and noncausal
filters. Again the performances are very similar. Moreover, after 45
dB, the rates are the same for both filters. Therefore, for the
ranges of parameters considered in these figures, causal filter
should be preferred over the noncausal one.

\begin{figure}
\begin{center}
\includegraphics[width = \figsize\textwidth]{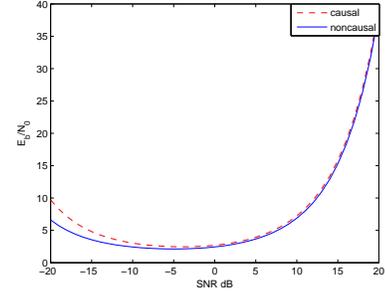}
\caption{Bit energy $\frac{E_{b}}{N_{0}}$ vs. SNR(dB) when
$\alpha=0.99$.} \label{fig:fig11}
\end{center}
\end{figure}

\begin{figure}
\begin{center}
\includegraphics[width = \figsize\textwidth]{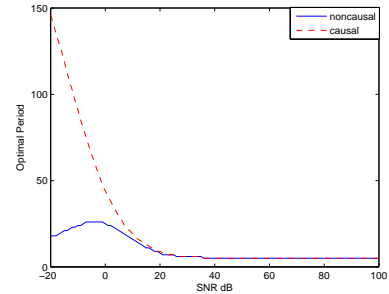}
\caption{Optimal period vs. SNR(dB) for causal and noncausal filters
when $\alpha=0.99$.} \label{fig:fig12}
\end{center}
\end{figure}

In systems where energy is at a premium, the energy required to send
one bit of information is a metric that can be adopted to measure
the efficiency of the system. The least amount of normalized bit
energy required for reliable communications is given by $
\frac{E_{b}}{N_{0}}=\frac{SNR}{C(SNR)} $ where $C(SNR)$ is the
channel capacity in bits/symbol. In our setting, we use the
achievable rates and analyze the required bit energy levels. In
Figure \ref{fig:fig11}, we plot the bit energy levels. The dashed
and solid lines show the results for causal and noncausal filters.
Note that the minimum bit energies are achieved  at SNR = -4dB and
-3dB for noncausal and causal filters, respectively. Operating below
these SNR levels should be avoided as it only increases the required
energy per bit. Figure \ref{fig:fig12} shows the optimal training
period values as a function of SNR for both filters. Interestingly,
the optimal period is increasing as SNR decreases for the causal
filter while it first increases and then decreases when the
noncausal filter is used. Since both past and future pilots are used
when a noncausal filter is employed, having large training periods
will diminish the benefits of future pilots especially for the data
symbols in the middle. Therefore, this option is avoided in this
case. On the other hand, having a larger period in the causal filter
case enables the system to put more power to the pilot by not using
data symbol slots farther away from the pilot and hence to obtain
more accurate channel estimates. In both filters, as SNR increases
the optimal period value stays constant at 5.

\section{Conclusion}
We have studied pilot-assisted communications when causal and
noncausal Wiener filters are employed at the receiver for channel
estimation. We have obtained achievable rate expressions by finding
the error variances in both cases. Subsequently, we have jointly
optimized the training period and power, and data power levels. We
have analyzed the effects of aliasing on the data rates in
Gauss-Markov Rayleigh fading channels when noncausal filters are
used. We have provided numerical results showing the optimal
parameters. We have compared the performances of causal and
noncausal Wiener filters at different SNR values. We have also
studied the energy-efficiency of pilot-assisted modulation with both
filters.

\end{document}